# VILEN MITROFANOVICH STRUTINSKY'S IMPACT ON NUCLEAR AND MANY PARTICLE PHYSICS

## Matthias Brack [1]

[1] *Institute for Theoretical Physics, University of Regensburg, D-93040 Regensburg, Germany*

**Abstract:** This paper is dedicated to the memory of Vilen Mitrofanovich Strutinsky who would have been 80 this year. His achievements in theoretical nuclear physics are briefly summarized. I discuss in more detail the most successful and far-reaching of them, namely (1) the shell-correction method and (2) the extension of Gutzwiller's semiclassical theory of shell structure and its application to finite fermionic systems, and mention some applications in other domains of physics.

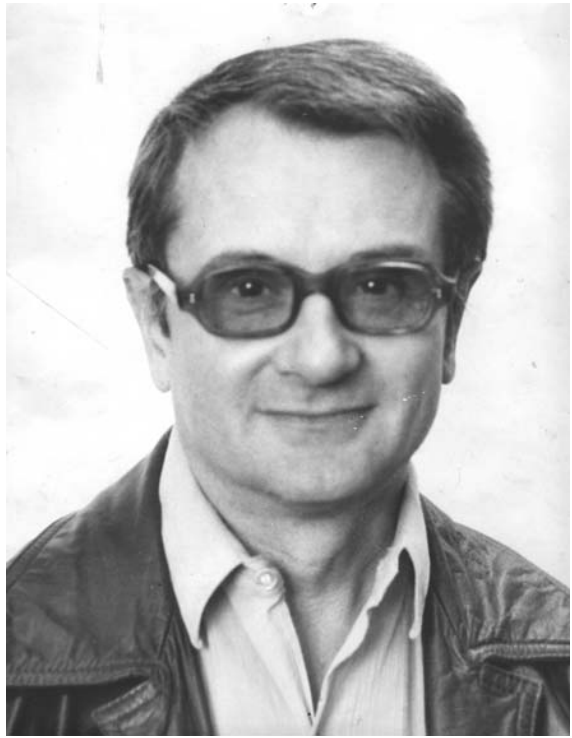

V.M. Strutinsky

### 1. Some personal recollections

*Copenhagen, 1968 - 1970*

My first contact with Vilen Strutinsky was made when I joined his research group at the Niels Bohr Institute (NBI) in Copenhagen in 1968, after having finished my diploma thesis at the University of Basel. It became my first experience in actual research during a time which may well be called the "big years of nuclear fission" at the NBI. In Strutinsky's group, large-scale calculations of nuclear deformation energies were performed using deformed Woods-Saxon potentials [1], employing his newly developed shell-correction method.

The results of this activity at the NBI were published in 1972 in a review article [2] which, upon Strutinsky's insistence (against the resistence of the editors), bears the title *"Funny Hills: The Shell-Correction Approach to Nuclear Shell Effects and Its Applications to the Fission Process"*. This title is characteristic of Strutinsky's humour which I came to like — in spite of the huge respect and fear which I initially had of him as a young student.

The following picture [3] was taken in 1972, on which he looks as I remember him from those years of scientific education and collaboration.

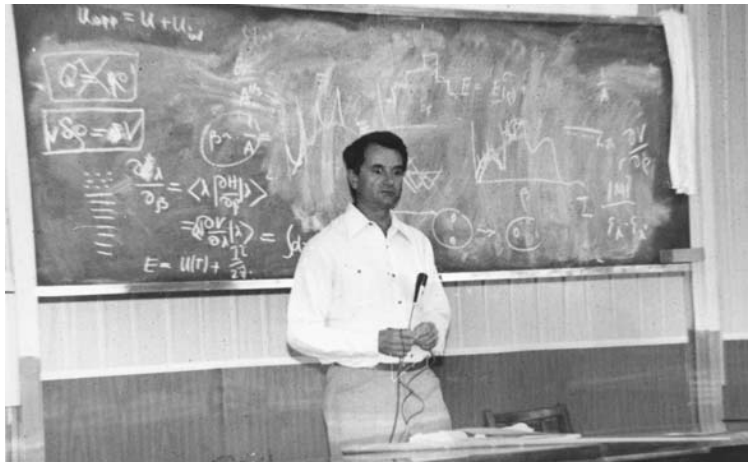

Report on the International School on Nuclear Structure, 1972

*Kiev, 1980 and 1982*

In the early '80s, when I was already established as a professor of physics at the University of Regensburg, I had the chance to visit Strutinsky in Kiev twice during about six weeks, both in 1980 and 1982. That time we worked on collective nuclear motion and started to develop the Liquid-Particle ("LiPa") model [4]. I came to know the members of his group at the Kiev Institute of Nuclear Research (KINR) and enjoyed the friendly and inspiring atmosphere here. Not to forget about the shahlik parties at the lake in the nearby forest!

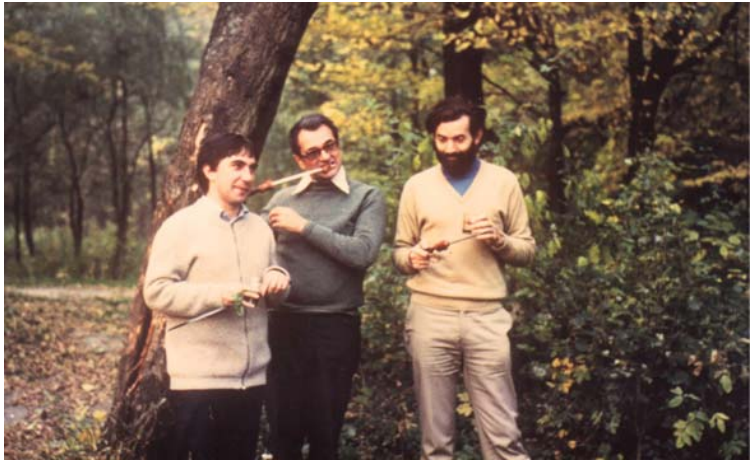

Party in the Golosejevo forest with guests of the Nuclear Theory Department,
C. Guet (left) and M. Brack (right), Kiev, October 1980

*Shorter meetings*

Besides those longer periods of collaboration, we met for shorter times at the NBI Copenhagen again in 1975 and 1989. In 1976, we met at a conference in Dubna. During the late '80s and early '90s, while he stayed for longer periods at Munich and Catania, Strutinsky visited my group at Regensburg University several times and, besides discussing physics, reminded me of the historical ties between the cities of Kiev and Regensburg.

He was my most important scientific mentor and teacher, and at the same time he became a dear friend. I owe to him the start of my scientific career, and a large part of my later research has been inspired by Strutinsky's ideas. He taught me not only physics, but also a lot about Ukrainian and Russian culture. We shared our passion for music. Over the years, I lost my initial fear of him, but I kept – and still keep – a deep respect for him.

Strutinsky was not an easy man. In public he could be aggressive and even offending — but in private he was one of the most considerate and compassionate humans I have ever met. He was an independent character in many respects. In science he had outstanding ideas which strongly influenced physics world wide, and he was ready to defend his ideas ferociously if needed. But he could also give in, when he accepted an opponent's arguments, and finish the dispute with a friendly gesture.

The following picture is characteristic of how many of us have known him:

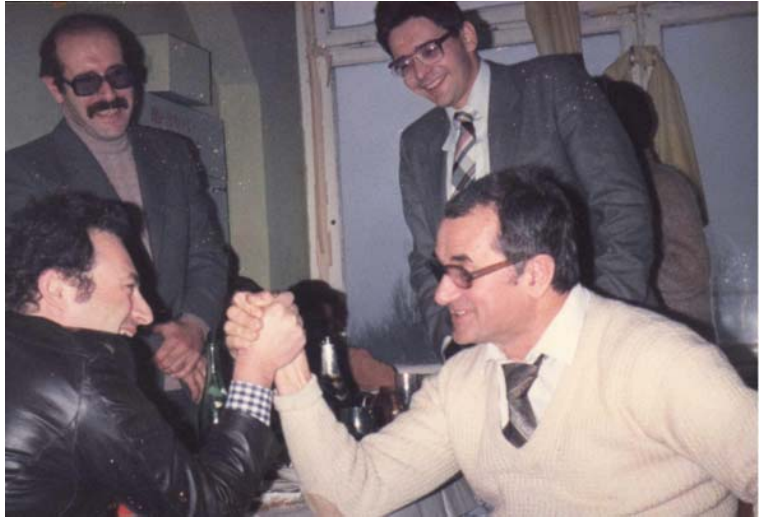

Sport competitions in the Nuclear Theory Department, 1988

Always ready for a fight — and a good joke at the end!

Here is a picture of Strutinsky the experimentalist, taken some time during the 1980s by V. Ostashko:

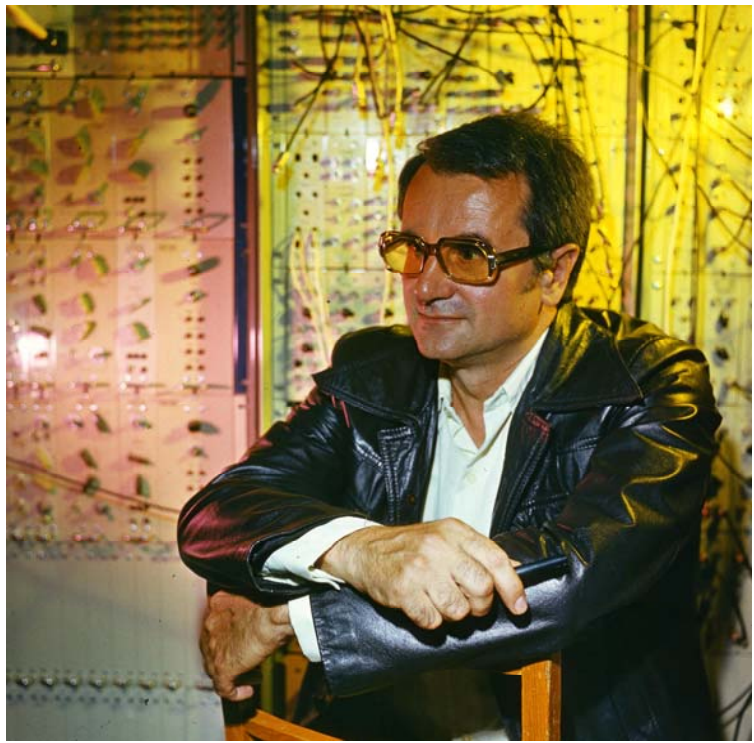

V.M. Strutinsky, the experimentalist (1980)

## 2. Strutinsky's scientific work

The Festschrift [3] published on the occasion of Strutinsky's 75th birth year quotes 139 published scientific articles. It would be impossible to render justice to his work in detail at this place. I will instead just sketch here some of its highlights, and discuss in some more detail those of Strutinsky's achievements which I deem to be the most pioneering and far-reaching for nuclear physics, and which had an important impact also on many particle physics in other domains: (1) the shell-correction method, and (2) the extension of Gutzwiller's semiclassical theory of shell structure and its application to finite fermionic systems. Since all the results that I have presented in my talk at this conference have been published, I shall not reproduce any of the figures here. Also, because this report cannot be made into a review article, I must refrain from giving references to well-known models and approaches, as well as to a huge body of literature on Strutinsky's shell-correction method in nuclear physics. I will quote mainly Strutinsky's own work, and work that is directly related to his primary ideas, as well as very recent extensions of his ideas and some suitable reviews.

*The early years*

Strutinsky's work during 1955 - 1965 dealt with various aspects of nuclear physics: decay modes, angular correlations, level densities, statistical theory of reactions, and the liquid drop model (LDM). Quite early, Strutinsky developed a particular interest in nuclear fission. With I. Halpern, he presented a paper on angular distributions of fission fragments at the UN Conference on Peaceful Uses Of Atomic Energy in Geneva 1957 [5]. At the IAEA Fission Conference in Vienna 1965, he presented papers on the fission of deformed nuclei [6] and on paring effects in fission [7]. In a skillful shape-independent variational calculation [8], he established the scission configuration quite generally within the LDM. This approach has recently been taken up again and mentioned at this conference [9].

*The shell-correction method*

In 1966, Strutinsky's first and certainly most far-reaching break-through occurred [10]. Until then, many attempts had been made to incorporate shell effects into nuclear deformation energies beyond the LDM. But they all failed in reproducing the fission barriers of actinide nuclei and details such as, e.g., the left-right asymmetry of the nascent fission fragments [11]. Summing the single-particle energies of a deformed shell model (like the Nilsson model) up to the Fermi energy failed at larger deformations. There was a need to renormalize the wrong average part of the single-particle energy sum. Knowing that the smooth part of the nuclear binding energy could be well described by the phenomenological LDM (or droplet models), Strutinsky wrote the total nuclear energy as

$$E_{tot} = E_{LDM} + \delta E, \qquad (1)$$

where $E_{LDM}$ is the LDM energy and $\delta E$ the so-called *"shell-correction energy"* which contains the fluctuating part of the single-particle energy sums for the neutrons,

$$\delta E = 2 \left( \sum_{n=1}^{N/2} E_n - < \sum_{n=1}^{N/2} E_n > \right), \qquad (2)$$

and similarly for the protons. Both parts of the total energy (1) depend on the neutron and proton numbers $N$ and $Z$, as well as on the nuclear deformation which has to be suitably parameterized both in the LDM and the shell model. While it had been a wide-spread belief that the shell-correction $\delta E$ was important only for spherical nuclei and would vanish at larger deformations, Strutinsky was convinced that shell effects play an important role also at larger deformations and lead, in fact, to new magic numbers corresponding to deformed systems with increased local stability [12].

For the determination of the smooth part $< \sum_{n=1}^{N/2} E_n >$ in (2), Strutinsky designed a very ingenious averaging method [10,12] which has been termed the *"Strutinsky smoothing method"*. It consists of a Gaussian convolution of the single-particle energy spectrum, modified (by the so-called *"curvature corrections"*) in such a way that the result does not depend on the energy averaging width $\gamma$ (at least within a finite interval of $\gamma$ of the order of the main shell spacing) and at the same time reproduces the true average level density which is found, e.g., from the extended Thomas-Fermi (ETF) model (or the Weyl expansion in the case of billiard or cavity models) (see, e.g., [13], Chap. 4, for details).

Strutinsky applied the shell-correction method to the calculation of fission barriers, employing the Nilsson model. For a typical actinide nucleus he obtained a second minimum in the deformation energy, lying above the ground state by about 3 MeV, at a deformation much larger than that of the ground state. This was, in fact, the physical explanation of the *fission isomer* which had been known experimentally but not understood theoretically. He presented this result at the Symposium *"Nuclides far off the stability line"* in Lysekil (Sweden) [14] in 1966, and he immediately became famous. He was then invited to the NBI in Copenhagen, in order to extend his calculations of fission barriers on a larger scale, employing more realistic shell-model potentials, which led to the team work published in *"Funny Hills"* [2].

The shell-correction method was soon taken up by many groups of the international scientific community. (On the other side of the Atlantic, it was mostly called the *"microscopic-macroscopic"* method.) Various combinations of nuclear shell models and liquid drop(let) models were used. Still today, the shell-correction method is being used world wide for calculations of nuclear masses and deformation energies, and it persists to yield the most accurate nuclear mass tables, ground-state deformations and fission barriers.

The energy averaging used for the second term in (2) is done in energy space and leads, as already mentioned, to the same results as the ETF model. However, it is formally not completely consistent with the particle-number averaging that is implicitly done in the standard least-square fits to the LDM which determines the smooth part of the total nuclear energy (1). Already early, alternative number averagings were investigated in [15]. Discrepancies in the resulting shell-correction energies remain, however, as discussed recently again [16, 17]. This point is thus still an object of current debate.

The decomposition of the total nuclear energy into a smooth and an oscillating part in (1) may at first glance look like a rather phenomenological ansatz. In particular, since the oscillating part is taken from the sum of occupied single-particle energies of the nucleons, one may argue that it cannot be correct, since this sum is well known not to represent the total energy in a self-consistent microscopic theory (where it double-counts the potential energy if the interaction does not depend on the density). However, it was soon realized that (1) is nevertheless correct even within a self-consistent microscopic theory. This was pointed out, amongst others, by Bethe [18] who therefore termed (1) the *"Strutinsky energy theorem"*. It can be rigorously proved that the oscillating part of the correct total energy is, indeed, contained in the single-particle sums; the proof is simply based on the variational principle that governs the self-consistent mean-field theories (see, e.e., [13], App. A.3), and it applies also to density-dependent nuclear interactions. This was soon demonstrated by the Strutinsky group [19] to hold also within the Migdal theory.

The Strutinsky energy theorem (1) was later tested numerically using the Hartree-Fock (HF) approach with effective Skyrme interactions. By extracting a self-consistent smooth part of the total HF energy (which is ideally represented by the LDM energy $E_{LDM}$), the first-order oscillating term was, indeed, found to be correctly represented by the shell-correction energy (2) evaluated in terms of the energy spectrum of the average self-consistent HF potentials (which are ideally represented by the shell-model potentials); the remaining terms were found to be less than ~ 1 MeV in all (not too small) nuclei [20].

The shell-correction method is thus a well-established practical approximation to a self-consistent microscopic theory and applicable to any bound many-fermion system. It has, e.g., been applied to metal clusters by many groups (see [21-24] for a few representative references). For semiconductor quantum dots, the Strutinsky energy theorem and shell-corrections to the Coulomb interaction energy were discussed in [25].

*The Kiev years*

Soon after his Copenhagen years 1967 - 1970, Strutinsky went to Kiev and formed his important research group at KINR. During the first years, efforts were made to theoretically consolidate the shell-correction approach on the level of many body theory, and to study alternative methods of averaging the single-particle energy spectrum, as already mentioned above. A huge activity of Strutinsky and his collaborators was, besides their continued interest in the interplay between shell effects and deformation, concerned with various aspects of nuclear dynamics (see [3] for references): e.g., large-amplitude collective motion (rotations and giant resonances), heavy-ion reactions and elastic scattering of heavy nuclei, dynamical aspects of fission, the dynamic extension of the shell-correction method termed "LiPa" model, developments of collective transport equations, a macroscopic generator coordinate method, etc. I cannot possibly do justice to all these activities in which I was not involved (apart from the "LiPa" model) and which lie beyond my competence. Many of them (and their further developments) have been presented in a number of talks at this conference.

In the following I want to present in some detail one activity of Strutinsky that had a large impact also on other domains of physics: the semiclassical description of shell effects.

*Semiclassical theory of shell structure*

In 1971, Gutzwiller published a semiclassical *"trace formula"* for the oscillating part of the single-particle level density of a Hamiltonian system [26,27], shortly to be followed by a similar formula applicable to billiard systems by Balian and Bloch [28] in 1972. Strutinsky discussed his energy averaging method at Saclay with Balian and Bloch, who had employed a similar energy averaging both to extract the smooth (Weyl) part of the level density of billiards [29] and to derive their trace formula, and thereby he became aware of semiclassical trace formulae and Gutzwiller's work.

The essence of a semiclassical trace formula is to relate the oscillating part of the level density of a quantum Hamiltonian system to the periodic orbits of the corresponding classical system. Dividing the quantum-mechanical level density into a smooth and an oscillating part:

$$g(E) = \sum_n \delta(E - E_n) = \tilde{g}(E) + \delta g(E), \tag{3}$$

the smooth part $\tilde{g}(E)$ can be determined in the ETF model (or by the Weyl expansion for billiards), while the oscillating part is given by the following *trace formula*:

$$\delta g(E) \simeq \sum_{PO} A_{PO}(E) \cos[S_{PO}(E)/\hbar - \sigma_{PO} \pi/2 - \varphi_d]. \tag{4}$$

The sum is over all periodic orbits (*PO*s) of the classical system, $S_{PO}(E) = \oint \mathbf{p} \cdot d\mathbf{q}$ are their action integrals, the amplitudes $A_{PO}(E)$ depend on their stabilities and degeneracies, and $\sigma_{PO}$ are called the Maslov indices. $\varphi_d$ is an extra phase that depends on the dimensionality of the system (and is zero when all orbits are isolated). The sum in (4) is an asymptotic one, correct to leading order in $1/\hbar$, and in non-integrable systems it is hampered by convergence problems [27]. For systems in which all orbits are isolated in phase space, Gutzwiller [26] expressed the amplitudes $A_{PO}(E)$ explicitly in terms of the periods and stability matrices of the *PO*s. His trace formula has become famous in particular in connection with *"quantum chaos"* [27].

The trace formula (4) thus relates the quantum oscillations in the level density to quantities that are determined purely by the classical system. Strutinsky, in his search for simple physical explanations of shell effects, realized that this kind of approach could help to understand the shell effects in terms of classical pictures. However, in the application to nuclear physics, Gutzwiller's expression for the amplitudes $A_{PO}(E)$ could not be used, because they diverge when the *PO*s are not isolated in phase space. This happens whenever a system has continuous (e.g., rotational) symmetries, and thus for typical shell-model potentials (except in nonaxially deformed situations). He therefore extended, with A. Magner, Gutzwiller's theory to systems with continuous symmetries [30,31]. (See also Magner's talk [32] and Chap. 6 of [13] for details.)

After that, trace formulae for systems with all kinds of mixed symmetries, including the integrable cases, were developed by various other authors. Uniform approximations were constructed for other situations where the original Gutzwiller theory could not be applied (orbit bifurcations and symmetry breaking under variation of the energy or a potential parameter); references to all these developments are given in [13], Chap. 6. The treatment of bifurcations is still an on-going subject of current research, see the talks of Magner [32] and Arita [33] who presented some beautiful examples of the role of bifurcations in connection with shell effects.

Strutinsky has, however, not only the merit of extending Gutzwiller's theory to realistic shell-model potentials, but he and his collaborators also extended the semiclassical approach to the description of bound many-fermion systems in the mean-field approach. In [30] it was shown that for such systems the shell-correction energy $\delta E$ (for one kind of particles) is given semiclassically by a similar-looking trace formula:

$$\delta E \simeq 2 \sum_{PO} A_{PO}(E_F)(\hbar/T_{PO})^2 \cos[S_{PO}(E_F)/\hbar - \sigma_{PO}\pi/2 - \varphi_d]. \qquad (5)$$

The difference to the trace formula (4) for the level density is that here the amplitudes and actions are to be evaluated at the Fermi energy $E_F$ of the considered kind of particles, and the appearance of an extra factor $(\hbar/T_{PO})^2$, where $T_{PO}$ are the periods of the $PO$s at the energy $E = E_F$. This extra factor brings a natural convergence to the sum, different from that in (4): orbits with longer periods contribute less to the shell-correction energy.

It is the beauty of this approach that gross-shell effects often can be explained in terms of a few of the shortest periodic orbits of a system. The earliest application of Strutinsky's group was given in [34], where the slopes of the stability valleys $\eta(N)$ of nuclei in a plot of $\delta E$ versus particle number $N$ and deformation $\eta$ were correctly reproduced using the shortest orbits in a deformed cavity (see also [31]). For other applications in various domains of physics (e.g., the mass asymmetry of the outer fission barrier, conductance oscillations in quantum dots and other mesoscopic systems, supershells in metal clusters, magnetization of quantum dots), I refer to Chap. 8 in [13]. Supershells in trapped cold fermionic atoms were also described semiclassically in [35], and shell effects in the moments of inertia were discussed in terms of periodic orbits in [36, 37]. Many further examples can be found in the literature – and will certainly also appear in the future!

In conclusion, it can be stated without exaggeration that Vilen Mitrofanovich Strutinsky has enriched physics by ingenious and beautiful ideas and very fruitful approaches. We are very grateful to him for that.

I thank Sasha Magner for a careful reading of the manuscript and its conversion into a "WORD" document.